\documentclass[aps,pre,showpacs,showkeys,twocolumn,superscriptaddress,
              floats,floatfix,preprintnumbers]{revtex4-1}
\usepackage{amssymb,amsmath}
\usepackage{bm,url,graphics,subfigure,epsfig,rotating,float,eufrak}
\usepackage{ulem}
\usepackage{color,soul}
\def \dtwo {d_{\rm 2}}
\def \PR {P_{\rm R}}
\def \Dtwo {D_{2}}
\def \zast {z^{\ast}}

\def \mP {\mathcal{P}}

\def \uu  {{\bm u}}
\def \vv  {{\bm v}}
\def \ff  {{\bm f}}

\def \ueta {u_{\rm \eta}}
\def \teta {t_{\rm \eta}}

\def  \XX  {{\bm X}}
\def \Xdot {\dot{{\bm X}}}
\def \vdot {\dot{{\bm v}}}
\def  \VV  {{\bm V}}

\def \p {p}
\def \cs {c_{\rm s}}
\def \Lx {L_{\rm x}}
\def \Ly {L_{\rm y}}
\def \Lz {L_{\rm z}}

\def  \RR  {{\bm R}}
\def  \VR  {V_{R}}
\def  \VRa  {|V_{R}|}
\def  \VRn  {V_{\rm n}}

\def   \zstar {z^\ast}

\def \taup {\tau_{\rm p}}

\def \curl {{\bm \nabla} \times}
\def \dive {{\bm \nabla}\cdot}

\def \delt {\partial_t}
\def \Dt {D_t}
\newcommand{\bra}[1]{\langle #1\rangle}
\def \Rey  {\mbox{Re}}
\def \Ma  {\mbox{Ma}}
\def \St  {\mbox{St}}

\def \Teddy {T_{\rm eddy}}

\def \kf  {k_{\rm f}}

\def \urms  {u_{\rm rms}}

\def \Np  {N_{\rm p}}

\def\d{{\rm d}}
\def\D{{\rm D}}

\def \pp {\mathfrak{p}}
\usepackage{pifont}
\newcommand{\I}{\mbox{\raisebox{-0.05cm}{\large\ding{192}}}}
\newcommand{\II}{\mbox{\raisebox{-0.05cm}{\large\ding{193}}}}

\newcommand{\fI}{\ensuremath{{{f_{1}}}}}

\newcommand{\Eq}[1]{Eq.~(\ref{#1})}
\newcommand{\Fig}[1]{Fig.~(\ref{#1})}

\newcommand{\bfig}{\begin{figure}}
\newcommand{\efig}{\end{figure}}
\newcommand{\bc}{\begin{center}}
\newcommand{\ec}{\end{center}}
\newcommand{\bea}{\begin{eqnarray}}
\newcommand{\eea}{\end{eqnarray}}

\begin{document}
\title{Statistics of the relative velocity of particles in turbulent
  flows : monodisperse particles}
\author{Akshay Bhatnagar}
\email{akshayphy@gmail.com}
\affiliation{ Nordita, KTH Royal Institute of Technology and
Stockholm University, Roslagstullsbacken 23, 10691 Stockholm, Sweden}
\author{K. Gustavsson}
\email{kristian.gustafsson@physics.gu.se}
\affiliation{ Department of Physics, Gothenburg University, 41296 Gothenburg, Sweden}
\author{Dhrubaditya Mitra}
\email{dhruba.mitra@gmail.com}
\affiliation{ Nordita, KTH Royal Institute of Technology and
Stockholm University, Roslagstullsbacken 23, 10691 Stockholm, Sweden}
\preprint{NORDITA 2017-108}
\begin{abstract}
We use direct numerical simulations to calculate the joint probability density function of
the relative distance $R$ and relative radial velocity component $\VR$ for a pair of heavy inertial particles suspended in
homogeneous and isotropic turbulent flows.
At small scales the distribution is scale invariant,  with a scaling
exponent that is related to the particle-particle correlation
dimension in phase space, $\Dtwo$.
It was argued~\cite{gustavsson2011distribution,gustavsson2014relative}
that the scale invariant part of the distribution has two asymptotic regimes:
{\I} $ \VRa \ll R$ where the distribution depends solely on $R$; and
{\II} $\VRa \gg R$ where the distribution is a function of $\VRa$ alone.
The probability distributions in these two regimes are matched along a straight line $\VRa = \zast R$.
Our simulations confirm that this is indeed correct.
We further obtain $\Dtwo$ and $\zast$ as a function of the Stokes
number, $\St$.
The former depends non-monotonically on $\St$ with a minimum at about
$\St\approx 0.7$ and the latter has only a weak
dependence on $\St$.
\end{abstract}
\maketitle
\section{Introduction}
\label{sec:intro}
Turbulent flows of gas with small heavy particles suspended in it is
at the heart many natural phenomenon; typical examples are:
(a) astrophysical dust is protoplanetary disks~\cite{Arm10},
(b) small water droplets in clouds~\cite{Pruppacher2010microphysics}, and
(c) aeolian processes (wind and sand)~\cite{kok2012physics}.
In all of these cases, usually two crucial questions are studied:
(a) whether the particles are homogeneously distributed in space or whether
they can form clusters? and (b) what is
the average collision velocities between the particles, which
in turn determines the collision kernel -- the number of collisions
per-unit-time, per-unit-volume.
Collisions between the particles play a crucial role in the
dynamics of these systems, for example, some collisions between the
water droplets in cloud may merge the droplets to form bigger
droplets. A similar process of collision and consequent merging of dust
grains plays a crucial role in formation of planetesimals (loosely
held dust balls of size of the order of kilometers) in protoplanetary
disks. Note that a complete knowledge of the collision kernel does not
allow us to determine the probability of merging or coagulation which
depends on one hand on the material properties of the colliding bodies
and on the other hand on the  probability distribution function of
collision velocities.
Clearly, a complete description of this problem is given by the joint
probability distribution function (JPDF) of the separation $\RR$
and relative velocity $\VV$ of a pair of particles.
The clustering of the particles in space and
the collision kernel can be calculated from the zero-th and first
moment of this JPDF respectively.
In its full generality this is an extremely difficult problem to solve.
Nevertheless, one can simplify the problem significantly and still
preserve the essence of it.
A commonly used model is that of heavy inertial particles, given by
\begin{subequations}
\begin{align}
\Xdot &= \vv \/, \label{eq:dxdt}\\
\vdot &= \frac{1}{\taup}\left[ \uu(\XX) - \vv \right] \/.
\label{eq:dvdt}
\end{align}
\label{eq:HIP}
\end{subequations}
Here the dot denotes time differentiation, $\XX$ and $\vv$ are
respectively the position and velocity of a particle,
$\taup$ is the characteristic relaxation time of the particle
and $\uu$ is the flow velocity that is determined by solving the
Navier--Stokes equation with appropriate boundary conditions.
Although the particles are approximated as points as far as the flow
is concerned, they are assigned finite sizes to calculate the collision
kernel.

The dynamics of heavy inertial particles in turbulence has been
extensively studied, starting with the pioneering work by
Saffman and Turner~\cite{saf+tur55}.
Since the turn of this century this problem has seen significant
progress. For a detailed description we direct the reader to several recent
reviews~\cite{sha03,gra+wan13,tos+bod09,gus+meh16,pumir2016collisional}.
Here we give a very brief summary of the results that are directly
relevant to this paper.
The equations of motion for heavy inertial particles, \Eq{eq:HIP},
are clearly dissipative even if the flow velocity is incompressible.
The stationary state of the system in phase-space is
characterized by an attractor with a correlation dimensions $D_2 < 2d$
where $d$ is the dimension of space~\cite{bec2003fractal}.
If $D_2 < d$ then the particles also show clustering in real
space~\cite{bec04} with a spatial correlation dimension $d_2 = D_2$.
This gives rise to small-scale clustering which has been studied
extensively using direct numerical simulations~\cite[see,
e.g.,][]{bec2007heavy}.
There are two fundamentally distinct mechanisms that brings particles
to close contact.
One is driven by the flow gradients~\cite{saf+tur55}, in which case the relative
velocities of the particles goes to zero as the particles smoothly
approach each other.
The second mechanism allows particles to have non-negligible relative
velocities at small distances~\cite{falkovich2002acceleration,wilkinson2006caustic}.
From \Eq{eq:HIP} it follows that under quite general assumptions the
gradient of the velocity of the HIPs,
$\sigma_{ij} \equiv \delta v_i/\delta X_j$
develops singularities, caustics, in
finite time~\cite{wilkinson2006caustic}.
This implies that the relative velocity of two particles does not go
to zero as their separation goes to zero but remains finite giving
rise to a high collision kernel.
Most of the theoretical, numerical, and experimental works have
concentrated on calculating the clustering exponent $d_2$ and the
collision kernel as a function of the Stokes number,
 $\St =\taup/\teta$,
where $\teta$ is the characteristic time at the dissipative scales.
The clustering exponent, $d_2$, can be obtained from the scaling
exponent of the zero-th moment of this JPDF and the collision kernel
is the first moment of the JPDF calculated at $\mid \RR \mid = 2 a$
where $a$ is the radius of a particle.
Our aim in this paper is to calculate
this JPDF, particularly its scaling behavior from direct
numerical simulations (DNS) of identical heavy inertial
particles in forced, homogeneous, and isotropic, turbulent flows.

A significant amount of
theoretical~\cite{gustavsson2011distribution, gustavsson2014relative},
experimental~\cite{de2010measurement, saw+bew+bod+ray+bec14}, and
numerical~\cite{sun+col97, wang2000statistical, hubbard2012turbulence,
 pan2013turbulence, per+jon15}
work has gone into studying the joint probability distribution
function, $\mP(\RR,\VV)$.
 Most of the numerical work has been limited to calculating $\mP$ for
few values of $R \equiv \mid \RR \mid$.
Some of the numerical works use either smooth random
flows~\cite{gustavsson2014relative,gustavsson2011distribution} or models
of turbulence~\cite[e.g., a shell model,][]{hubbard2012turbulence}
instead of direct numerical simulations (DNS).
The earliest DNS~\cite{sun+col97} already pointed out that $\mP$ for
a fixed $R$  is not Gaussian, but possesses exponential tails.
Some of the later studies~\cite{wang2000statistical, hubbard2012turbulence,
pan2013turbulence} have confirmed the exponential tails and have
demonstrated clear asymmetry between positive -- a pair of particles
moving away from each other -- and negative -- a pair of particles
moving towards each other -- side of $\mP$.
Ref.~\cite{gustavsson2011distribution} is the first paper to write
down the Fokker-Planck equations satisfied by $\mP$ by virtue of using
a one-dimensional, random, smooth, white-in-time model for the flow
velocity.
The scaling behavior of $\mP$ is obtained by solving the Fokker-Planck
equation.
Guided by this model, Ref~\cite{gustavsson2014relative} has argued
that the JPDF, $\mP$, possesses certain symmetries in quite general
cases.
A recent paper~\cite{per+jon15} has used DNS to study the scaling
behavior of the JPDF, $\mP$, for several different values of $R$ and have
confirmed some of the conclusions of
Ref.~\cite{gustavsson2014relative} for $\St \approx 1$.
Our aim in this paper is to calculate $\mP$, particularly its
scaling behavior, from direct numerical simulations of identical
heavy  inertial particles in forced, homogeneous, and isotropic, turbulent
flows.

The rest of this paper is organized as follows:
in section~\ref{prelim}
we briefly recapitulate the main theoretical results
on scaling properties of $\mP$, followed by section~\ref{model} where
we describe in detail our direct numerical simulation;
in section~\ref{results} we show that our DNS
indeed confirms these theoretical results, in particular, the
symmetries and the scaling nature of $\mP$; we conclude in section~\ref{conc}.

\section{Theoretical Background}
\label{prelim}
\begin{figure}
\begin{center}
\includegraphics[width=0.8\columnwidth]{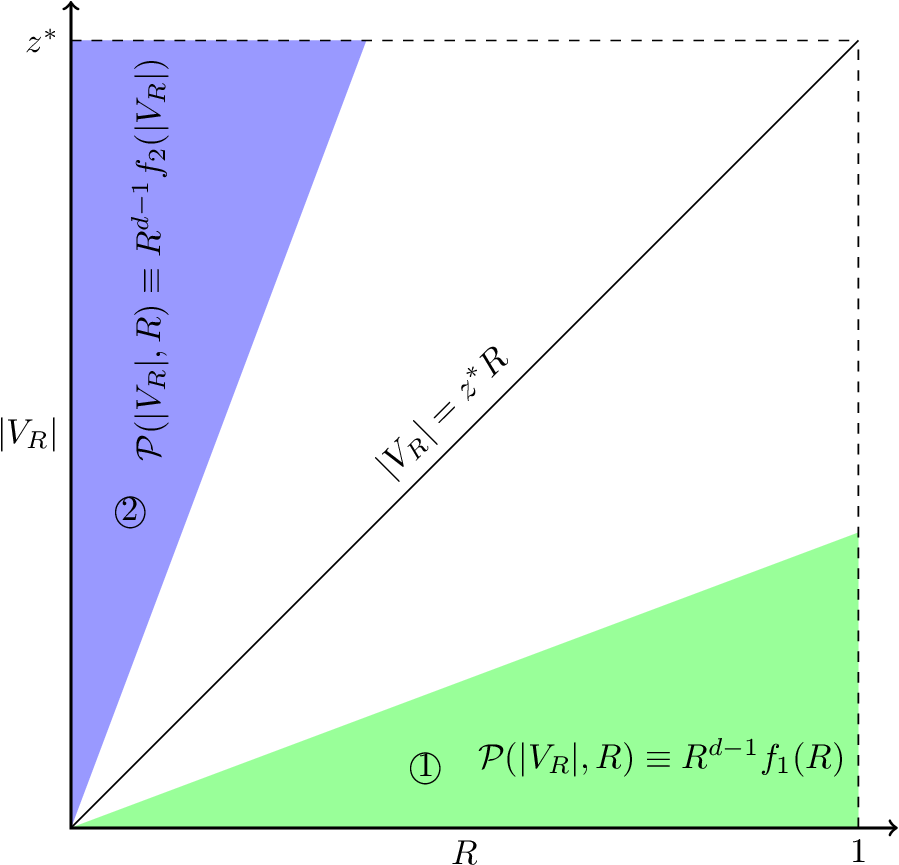}
\caption{
Schematic diagram of $R-\VRa$ phase space. Light blue
  (\II) region of the phase space corresponds to the region where $\VRa \gg R$.
In this region the contribution due to caustics dominates and the
joint probability density function $\mP(R,\VRa)$ becomes
independent of $R$.
The light green region ({\I}) corresponds to the region where $\VRa \ll R$.
In this region the relative velocity of the particles $\VR$ is a smooth function
of their separation $R$, and  $\mP(R,\VRa)$ is only a function
of $R$. Two functions are matched along the line $\VRa=\zstar R$ to obtain the $\mP(R,\VRa)$.
}
\label{fig:rvspace}
\end{center}
\end{figure}
We consider the flow to be forced at large length scales,
statistically stationary, homogeneous and isotropic.
The model of heavy inertial particles, \Eq{eq:HIP},
is applicable if the size of the particles are smaller than the
smallest energy carrying length scale of the turbulence -- the Kolmogorov scale,
$\eta$. The characteristic velocity at the length scale is $\ueta$
such that the scale-dependent Reynolds number of the
Kolmogorov scale is $\Rey_{\eta} = \ueta\eta/\nu=1$ where $\nu$ is the
kinematic viscosity of the fluid. As we are primarily interested
in particle collisions, we are interested
in relative velocities of particles at small separations
-- smaller than $\eta$. In what follows we use $\eta$ and
$\ueta$ as our characteristic scales of length and velocity
respectively.

As mentioned in the introduction two fundamentally distinct mechanisms
may bring particles into contact at small separations in turbulent aerosols.
In the first case~\cite{saf+tur55} the particles are brought together
by the turbulent flow velocity; they spend a long time together, smoothly
approaching each other towards small spatial separations.
Their phase-space separation approaches zero as their relative distance goes to zero.
The second
possibility~\cite{falkovich2002acceleration,wilkinson2006caustic,gustavsson2014relative}
is that the particles detach from
the flow, allowing caustics to form, leading to a multi-valued particle velocity field.
If the detachment is large enough compared to the distance between two
particles they may move towards each other close to ballistically.
The latter effect give rise to particle collisions with large relative
velocities -- the phase-space separation remains finite as the distance approaches zero.
These observations were used in
Refs.~\cite{gustavsson2011distribution,gustavsson2014relative}
to calculate the asymptotic behavior of the joint probability
distribution of separations and relative velocities, $\mP(\RR,\VV)$,
in a smooth, homogeneous and isotropic flow.

The radial projection
$\VR = (\VV \cdot \RR)/R$ of $\VV$ plays a crucial role
in the collision between the particles.
As two particles detach from the relative flow velocity
at large separations, their ballistic motion will bring them
into contact at small separations if their tangential velocity is
close to zero, i.e. if $\VV\approx \VR\RR/R$ and if $\VR$ is negative.
In spherical coordinates, in $d$ dimensions, we write the joint
distribution of $R$ and $\VRa$ as
\begin{equation}
\mP(R,\VRa) = R^{d-1}f(R,\VRa)\,,
\label{eq:P_R_Vr_GeneralForm}
\end{equation}
where the factor of $R^{d-1}$ comes from the spatial volume element.
This volume element corrects for the probability of having a small
tangential velocity, and we can think of $f(R,\VRa)$ to be the
probability to have velocity $\VRa$ at distance $R$ in the
two-dimensional system formed by $R$ and $\VRa$.
We summarize the theoretical
predictions~\cite{gustavsson2014relative} of $\mP(R,\VRa)$ below.
\begin{enumerate}
\item \label{mirror}
The distribution $\mP$, reduces to two asymptotic forms
\begin{align}
\label{eq:PVr_ansatz}
P(R&,\VRa)\sim
\\
&R^{d-1}
\left\{
\begin{array}{lll}
\fI(R,\VRa) & \mbox{for }\VRa\le \Upsilon(R) & \I\cr
\fI(\Upsilon^{-1}(\VRa),\VRa) & \mbox{for }\VRa>\Upsilon(R) & \II\cr
\end{array}
\right.\,.
\nonumber
\end{align}
These two asymptotic limits are illustrated as region {\II} and region
{\I} respectively in \Fig{fig:rvspace}, and are matched along an
curve  $\VRa=\Upsilon(R)$ [The curve $\Upsilon(R)=\zstar R$ is shown in \Fig{fig:rvspace}].
\item \label{zast}For relative distances smaller than the
  dissipation scale, i.e., $R < 1$, the flow velocity can be
  approximated by a smooth flow, in which case the  matching curve is
  found to be
\begin{equation}
\Upsilon(R) = \zast R \quad \mbox{for} \quad R < 1 \/,
\label{eq:match}
\end{equation}
where $\zast$ is a constant.
\item \label{fscaling}The scaling function in the dissipation range is,
\begin{equation}
\fI(R,\VRa) \sim R^{D_2-d-1} \quad \mbox{for} \quad R < 1 \,,
\label{eq:fscaling}
\end{equation}
where $\Dtwo$ is the correlation dimension of the attractor
of the long-time stationary state of the particles in phase space.
\end{enumerate}
We shall call the predictions \ref{mirror} and \ref{zast} together
the ``asymptotic mirror symmetry'' of $\mP$ in the dissipation range. This is because
the distribution in region {\I} is effectively mirrored in the curve $\VRa=\zstar R$, giving the 
distribution in region {\II}.
In this paper we show, from direct numerical simulations, that
all the above predictions hold.

\section{Direct Numerical Simulations}
\label{model}
The flow velocity $\uu$ is obtained by
direct numerical simulation (DNS) of the Navier--Stokes
equation,
\begin{subequations}
\begin{align}
\partial_t \rho &+ \nabla \cdot (\rho \uu) = 0  \/, \label{eq:density}\\
\rho \Dt \uu &= -\nabla \p + \mu \nabla \cdot {\bm S} + {\bm f}  \/,
\label{eq:mom}
\end{align}
\label{eq:fluid}
\end{subequations}
under isothermal conditions, with an external force.
Here $\Dt \equiv \delt + \uu \cdot \nabla$ is the advective
derivative,  $\p$, and $\rho$ are
respectively the velocity, pressure, and density of the flow,
$\mu$ is the dynamic viscosity, and
${\bm S}$ is a second-rank tensor with components
$S_{kj} \equiv \partial_k u_j + \partial_j u_k -\delta_{jk}(2/3)\partial_k u_k$.
The simulations are performed in a three-dimensional periodic box
with sides $\Lx=\Ly=\Lz=2\pi$.
In addition we use the ideal gas equation of state with a constant
speed of sound $\cs = 1$.

We use the Pencil-Code~\cite{pencil-code},
which uses a sixth-order finite-difference scheme for space
derivatives and a third-order Williamson-Runge-Kutta~\cite{wil80} scheme for time
derivatives.
The external force ${\bm f}$ is a white-in-time stochastic
process  that is integrated by using the Euler--Marayuma scheme~\citep{hig01}.
The same setup have been used in
studies of scaling and intermittency  in fluid and magnetohydrodynamic
turbulence~\cite{dob+hau+you+bra03,hau+bra+dob03,hau+bra04}.
We introduce the particles into the DNS
after the flow has reached statistically stationary state.
Then we simultaneously solve the equations of the flow, \Eq{eq:fluid},
and the heavy inertial particles, \Eq{eq:HIP}.
To solve for the particles in the flow we have to interpolate the flow velocity to
typically off-grid positions of the particles. We use a tri-linear method for
interpolation.

The flow attains a statistically stationary state when
the average energy dissipation by viscous forces is balanced by
the average energy injection by the external force $\ff$ which is
 concentrated on a shell of wavenumber with
radius $\kf$ in Fourier space~\cite{B01}.
We define the Reynolds number by $\Rey\equiv \urms/(\nu\kf)$,
where $\urms$ is the root-mean-square velocity of the flow
averaged over the whole domain and the kinematic
viscosity $\nu =\mu/\bra{\rho}$ where the symbol
 $\bra{\cdot}$ denotes spatial and temporal averaging over the
statistically stationary state of the flow.
The mean energy dissipation rate is
$\varepsilon \equiv 2\nu \bra{\omega^2}$, where
$\omega \equiv \curl \uu$ is the vorticity.
The Kolmogorov scale or the dissipation length scale is
given by
$\eta \equiv (\nu^3/\varepsilon)^{1/4}$,
the characteristic time scale of dissipation is given by
$\teta = \sqrt{\nu/\varepsilon}$
and
$\ueta \equiv \eta/\teta$ is the characteristic velocity scale
at the dissipation length scale.
In our simulations, unless otherwise stated, we use
$\eta$, $\teta$, and $\ueta$ to non-dimensionalize
length, time, and velocity respectively.
The large eddy turnover time is given by
$\Teddy\equiv 1/\kf\urms$.
We define the Stokes number $\St \equiv \taup/\teta$.

The amplitude of the external force is chosen such that the
Mach number, $\Ma \equiv \urms/\cs$ is always less than $0.1$, i.e.,
the flow is weakly compressible.
If on one hand we consider application of our DNS to understand
rain-formation in Earth's atmosphere it would be appropriate to
consider an incompressible flow.
We have checked that the weak compressibility in our DNS does not
have any significant effect  in the following manner:
(a) The probability density of $\dive \uu$ along particle trajectories is
found to be a Gaussian highly peaked at zero,
see Appendix~\ref{app:divu}, \Fig{fig:divu}.
(b) We parametrise the compressibility by calculating the dimensionless number
$\pp=\bra{\mid\dive \uu\mid^2}/\bra{\mid\curl\uu\mid^2} = 8.6\times 10^{-3}$.
Following Ref.~\cite{gus+meh16} we estimate that $\pp=8.6\times 10^{-3}$
implies that
the compressibility of the flow may have significant effect only on particles with
$\St \lesssim 0.11$ whereas the smallest value of $\St$ used in our simulations is $0.17$.
(c) And finally we calculate the correlation dimension, $\dtwo$, of the cluster formed by the
particles to find that our results agree with those obtained in
incompressible flows~\cite{bec2007heavy}.
On the other hand, if we consider possible application of our work to
astrophysical problems, e.g., to protoplanetary
disks~\cite{pan2013turbulence} then indeed the weakly compressible
simulations are the appropriate model.
\begin{table*}
\begin{center}
\caption{Table of parameters for our DNS runs with $N^3$ collocation points: $\nu$ the coefficient of
kinematic viscosity, $\Np$ is the number of particles,
$\epsilon$ in the mean rate of energy dissipation, $\eta \equiv
(\nu^3/\epsilon)^{1/4}$
and $\teta = \sqrt{(\nu/\epsilon)}$ are the dissipation length and time
scales, respectively, $\Rey$ is the Reynolds
number based on forcing length scale, and $\Teddy$ is large eddy turn-over time scale of the flow.}
\begin{tabular}{c c c c c c c c c}
\hline\hline
Run & $N$ & $\nu$ & $\Np$ & $\Rey$ & $\varepsilon$ & $\eta$ & $\teta$ & $\Teddy$ \\
\hline
{\tt R1} & $256$ & $1.0\times 10^{-3}$ & $10^7$ & $41$ & $3.02\times 10^{-3}$ & $2.4\times
10^{-2}$ & $0.56$ & $0.93$ \\
{\tt R2} & $512$ & $5.0\times 10^{-4}$ & $10^7$ & $89$ & $3.25\times 10^{-3}$ & $1.4\times
10^{-2}$ & $0.39$ & $0.86$ \\
\hline
\end{tabular}
\end{center}
\label{table:para}
\end{table*}

\section{Results}
\label{results}

To calculate the JPDF we need to look at all pairs of particles
within a separation of $\eta$.
Naively speaking this process scales quadratically with the number of particles, $\Np^2$, but using the standard
technique of constructing linked-lists -- commonly used in molecular
dynamics  simulations~\cite[see, e.g.,][Chapter 5]{All+Til89},
we can reduce number of computations to be proportional to $\Np$.
The same technique has been used before in
Refs.~\cite{sun+col97,per+jon15}.
\subsection{Correlation dimensions}
\label{subsec:cluster}
\begin{figure}
\begin{center}
\includegraphics[width=0.9\columnwidth]{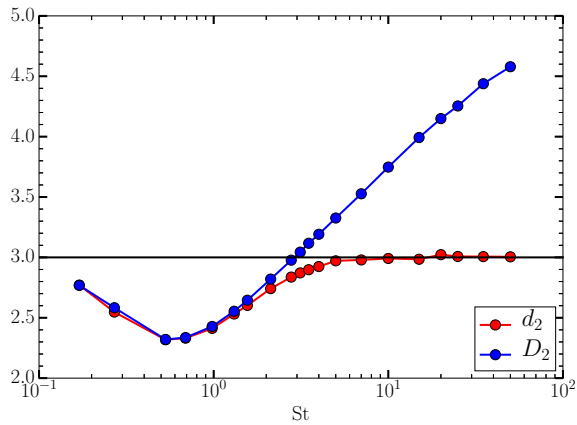}
\caption{
\label{fig:D2}
Phase-space correlation dimension $\D_2$ and real-space
correlation dimension $\d_2$ as functions of $\St$. Black dashed
curve shows the data from Ref~\cite{bec2007heavy}.}
\end{center}
\end{figure}
To calculate the phase-space correlation dimension we
evaluate the $\mP(w)$ which is the probability distribution function
(PDF) of $w$, where $w^2 = \mid \RR \mid^2 +\mid \VV\mid^2$.
We calculate $\Dtwo$ by the scaling behavior:
\begin{equation}
\mP(w)\sim w^{D_2-1} \quad {\rm as} \quad w\to 0
\end{equation}

In \Fig{fig:D2} we plot $\Dtwo$ as a function of $\St$
for the run {\tt R2}.
In the range of $\St$ we have calculated, $D_2$ is non-monotonic
function of $\St$ with a minimum near $\St \approx 0.7$
and does not change significantly as a function of $\Rey$.
Next we calculate the real-space radial distribution function (RDF), $g(R)$,
which is the PDF of $R$.
At small $R$, the RDF show scaling behavior:
\begin{equation}
g(R) \sim R^{d_2-1} \quad {\rm as} \quad R\to 0
\label{eq:d2}
\end{equation}
The small scale clustering of HIPs in real space is parametrised by
$d_2$ which is plotted for the
run {\tt R2}.
The values of $\dtwo$ we obtain is equal to (within error bars) the
earlier value of  $\dtwo$ obtained in Ref~\cite{bec2007heavy}.
Note that a crucial component of the theory described in
Ref.~\cite{gustavsson2014relative} is not $\dtwo$ but
$\Dtwo$ -- the phase-space correlation dimension.
Under fairly general conditions, it can be shown that~\cite{bec04} if
the particles cluster on a fractal set in phase-space with correlation
dimension  $\Dtwo$ then their real space correlation dimension
$\dtwo = {\rm min}( \Dtwo,d)$ where $d=3$ is the dimension of space.
To the best of our knowledge, $\Dtwo$ has never been calculated
before for HIPs in turbulent flows, although they have been calculated
for smooth random flows.
\subsection{Joint PDFs of $R$ and $\VRa$}
\label{subsec:jpdf}
\begin{figure*}
\begin{center}
\includegraphics[width=0.48\linewidth]{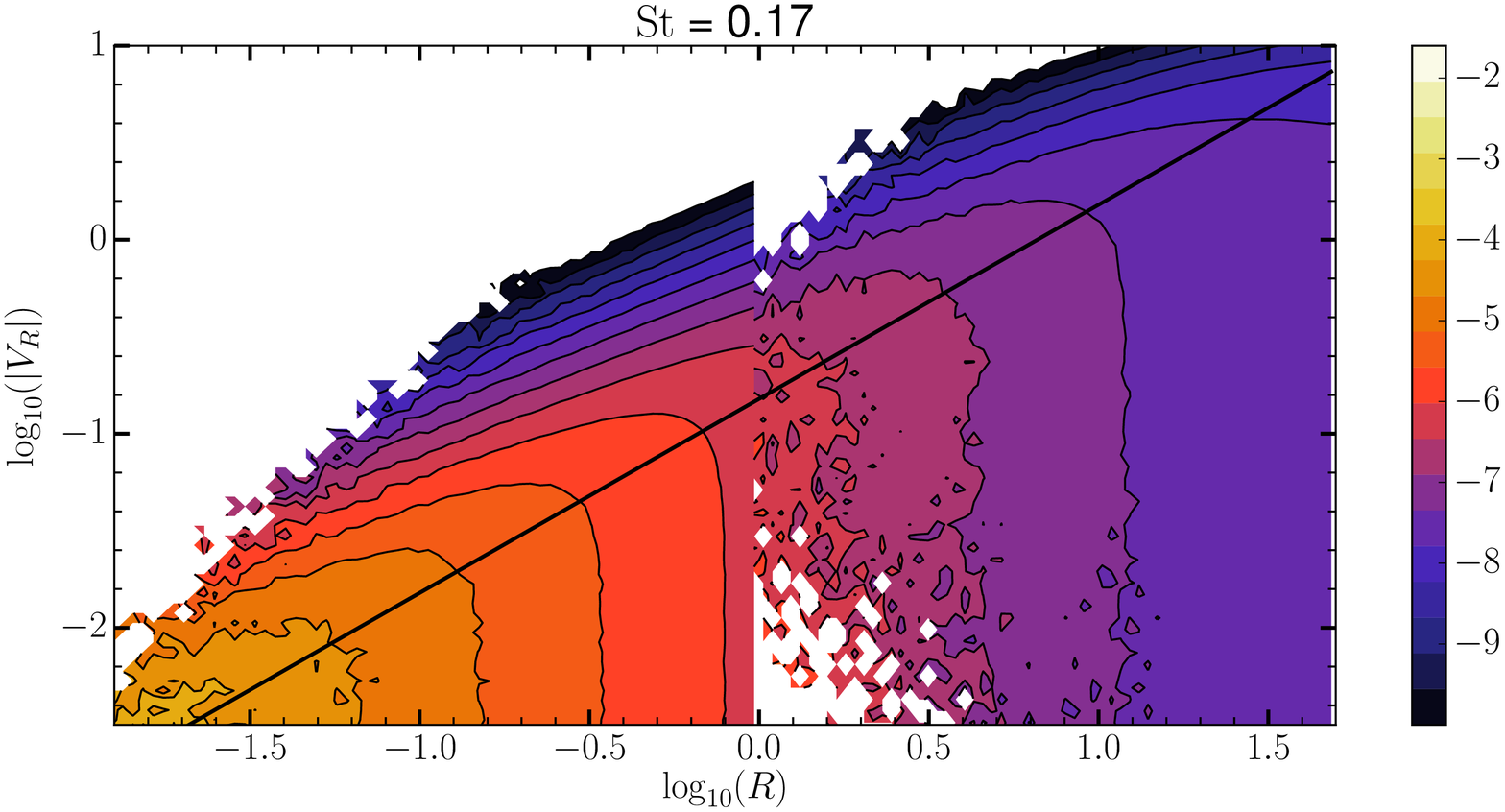}
\put(-225,100){{\bf (a)}}
\includegraphics[width=0.48\linewidth]{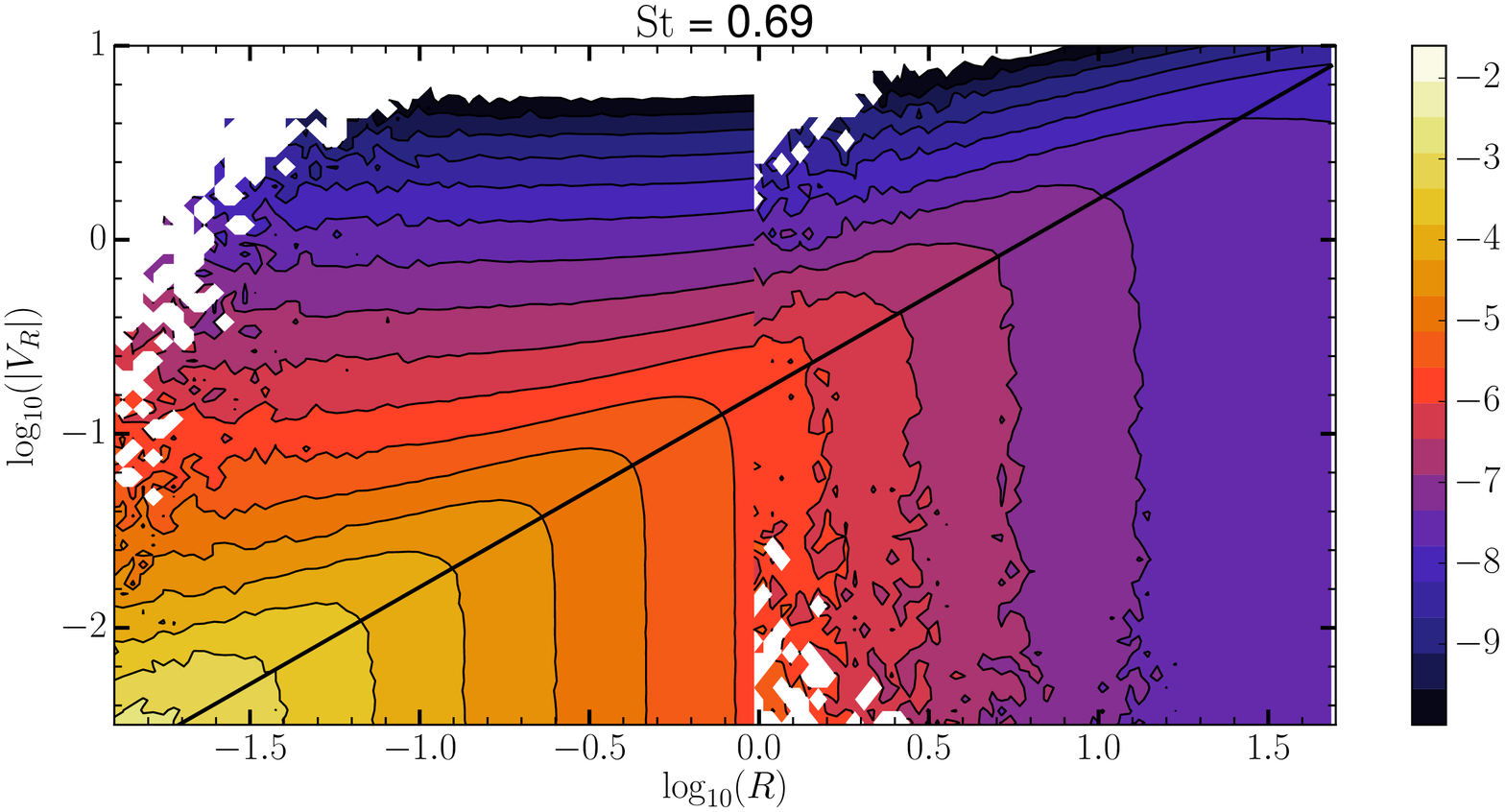}
\put(-225,100){{\bf (b)}}\\
\includegraphics[width=0.48\linewidth]{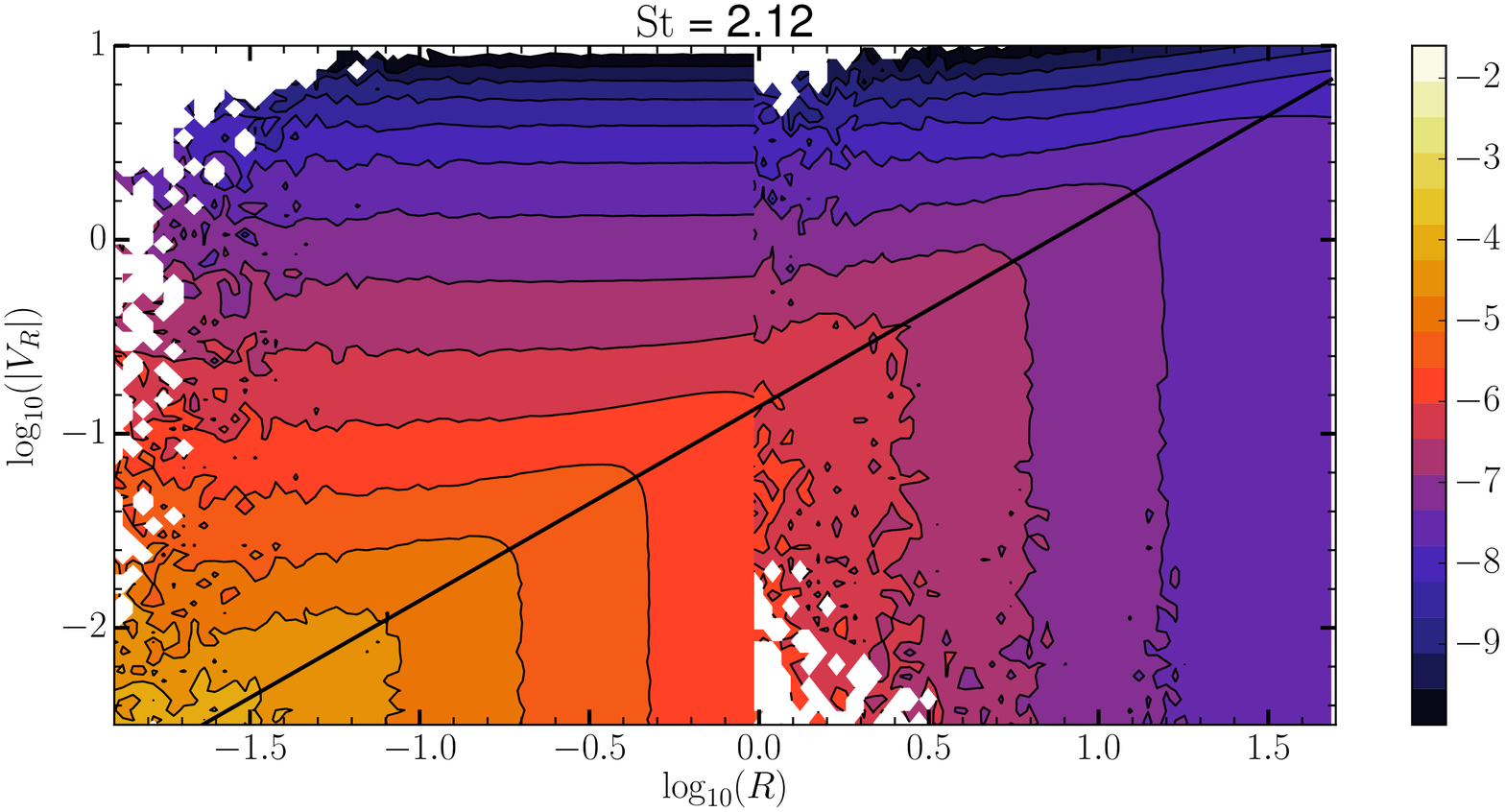}
\put(-230,107){{\bf (c)}}
\includegraphics[width=0.48\linewidth]{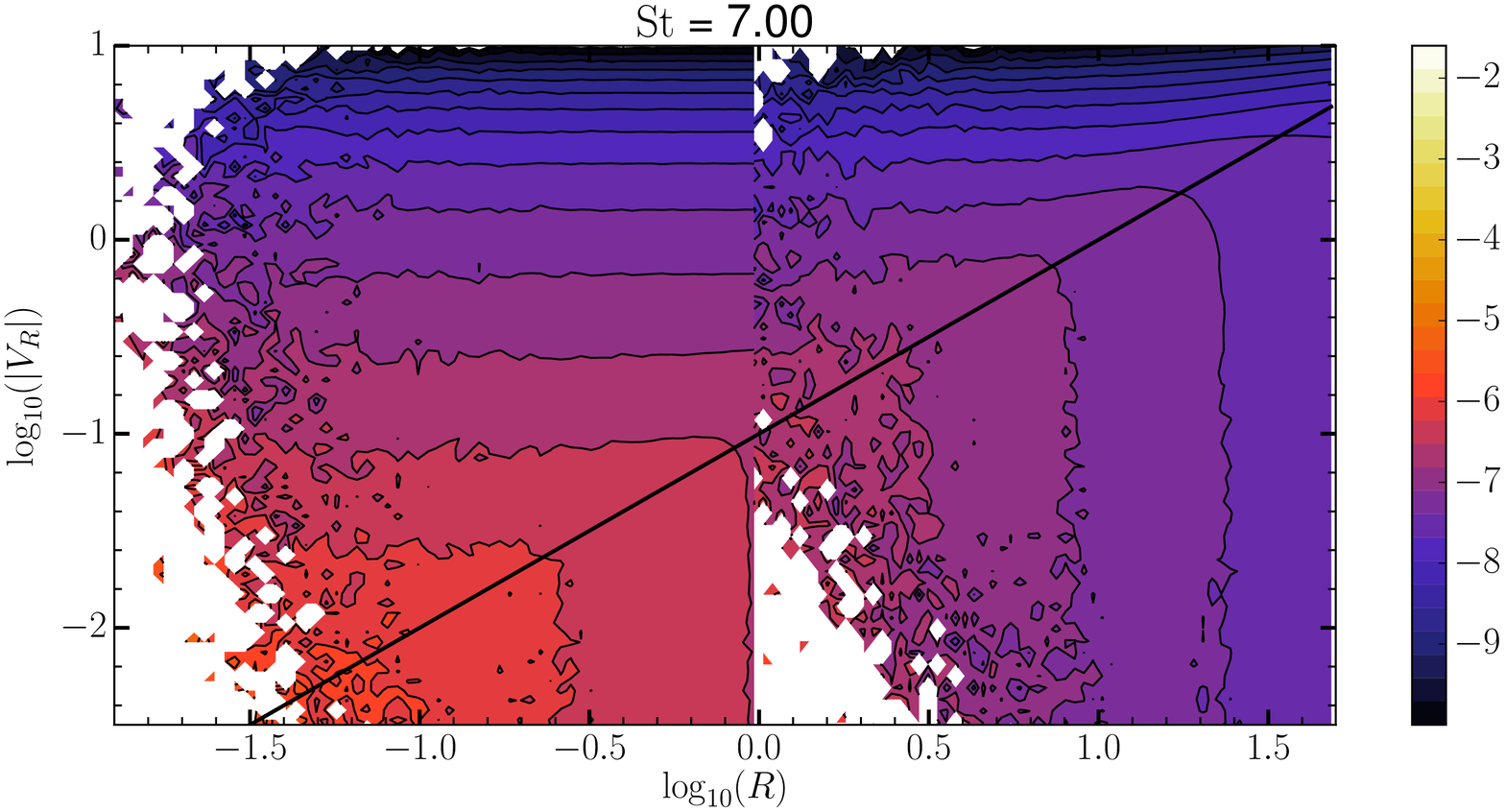}
\put(-230,107){{\bf (d)}}
\caption{(color online) Contour plots of joint PDFs, $\mP(R,\VRa)$,
  divided by $R^2$, for four representative values of $\St$ plotted in
  logarithmic scales. The joint PDF for $R<1$ and $R>1$ are calculated
  separately and then patched together.
}
\label{fig:jpdf}
\end{center}
\end{figure*}
Before we present our results on the JPDF, $\mP(R,\VRa)$ let us note
that the scaling theory of Ref.~\cite{gustavsson2011distribution} does not distinguish between
the positive and negative components of $\VR$, i.e., it does not
distinguish between two particles approaching each other and moving
away from each other.
In the same spirit, unless otherwise stated, we present below the
numerical data on $\mP(R,\VRa )$.

We plot in  \Fig{fig:jpdf} contour plots of
the joint PDF, $\mP(R,\VRa)$ for four different values of $\St$.
Let us first consider the top figures in the left column, $\St = 0.17$.
Looking at the region {\I} of these figures we find that
the contour lines are vertical, i.e., in region {\I}
$\mP(R,\VRa)$ is a function of $R$ alone.
But we have practically no data in the region corresponding to
{\II} in this figure.
This is expected because in region {\II} the contribution due to caustics dominate $\mP$, but caustics are exponentially suppressed with $\St$ for small values of $\St$~\cite{wilkinson2006caustic}.
Note also that for a fixed but small $\St$ we should always be able to
find the contribution from the caustics if we can probe small enough
$R$.
But at small $\St$ there are few particle pairs whose separation is
very small. This is because the real-space clustering exponent $d_2$ is
quite close to $d=3$ for small $R$.
These two factors explain why we have difficulty observing the
contribution from the caustics at small $\St$.
Looking at the higher values of $\St$ in
\Fig{fig:jpdf}, we indeed find that the contour lines of $\mP$ in
region {\II} become horizontal,
i.e., $\mP(R,\VRa)$ becomes a function of $\VRa$ alone.
We estimate the matching scale $\zast$ from left column of
\Fig{fig:jpdf} by fitting a line through the points where contours
change from vertical to horizontal.
This line is the matching curve $\Upsilon(R)$.
It is clear from the figure that this matching line continues beyond
its theoretically predicted regime, $R<1$ to at least up to
$R \gtrsim 10$
for all the $\St$ values we have studied.

\begin{figure}
\begin{center}
\includegraphics[width=0.9\linewidth]{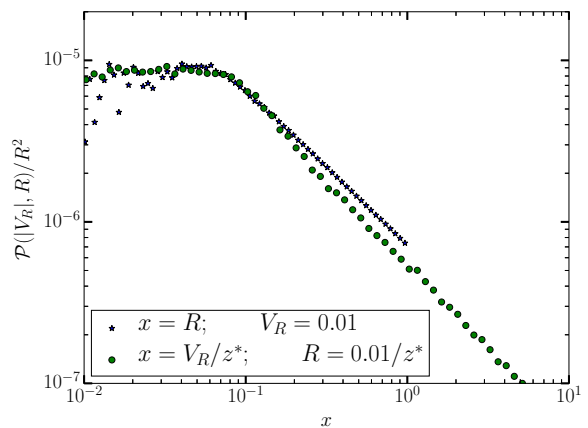}
\caption{(color online) The joint PDF, $\mP$, for $\St = 3.13$,
as a function of $R$ for a fixed $\VRa$ plotted with symbols $\ast$. The same PDF
plotted along a vertical line, which is the mirror image of the
horizontal line about the line $\Upsilon(R) = \zast R$
is plotted with the symbol $\cdot$.
}
\label{fig:rvr}
\end{center}
\end{figure}
Next, to test the asymptotic mirror symmetry of $\mP$, we
plot in \Fig{fig:rvr},  $\mP$ along a horizontal line
as a function of $R$, for $R<1$, for a small value of $\VRa$.
Then we reflect this line about the matching curve $\Upsilon(R)$
to obtain a vertical line and then plot $\mP$ along this vertical line
in the same plot,~\Fig{fig:rvr}.
Thus we confirm the asymptotic mirror symmetry of $\mP$.

\begin{figure}
\begin{center}
\includegraphics[width=0.85\linewidth]{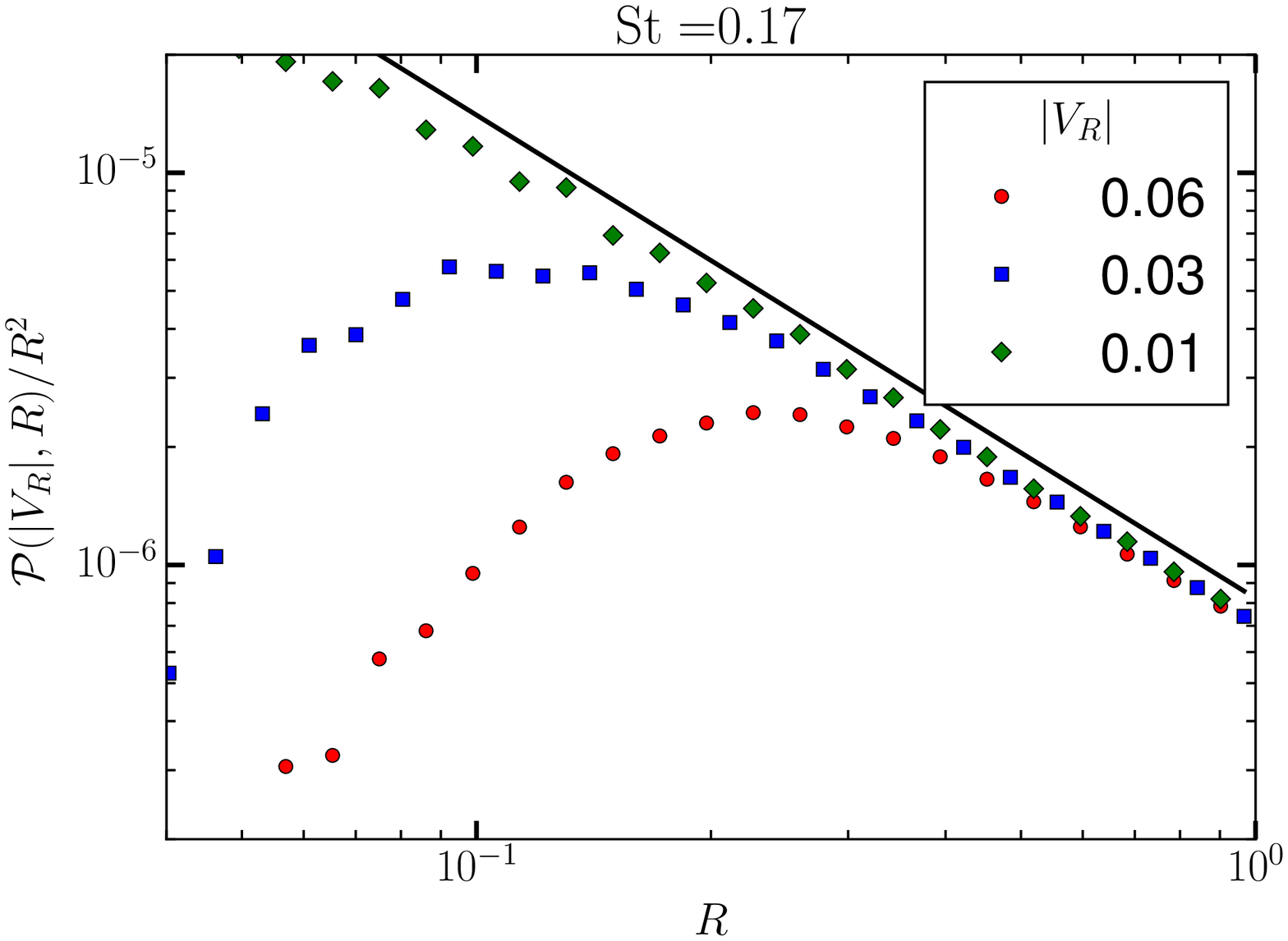}\\
\includegraphics[width=0.85\linewidth]{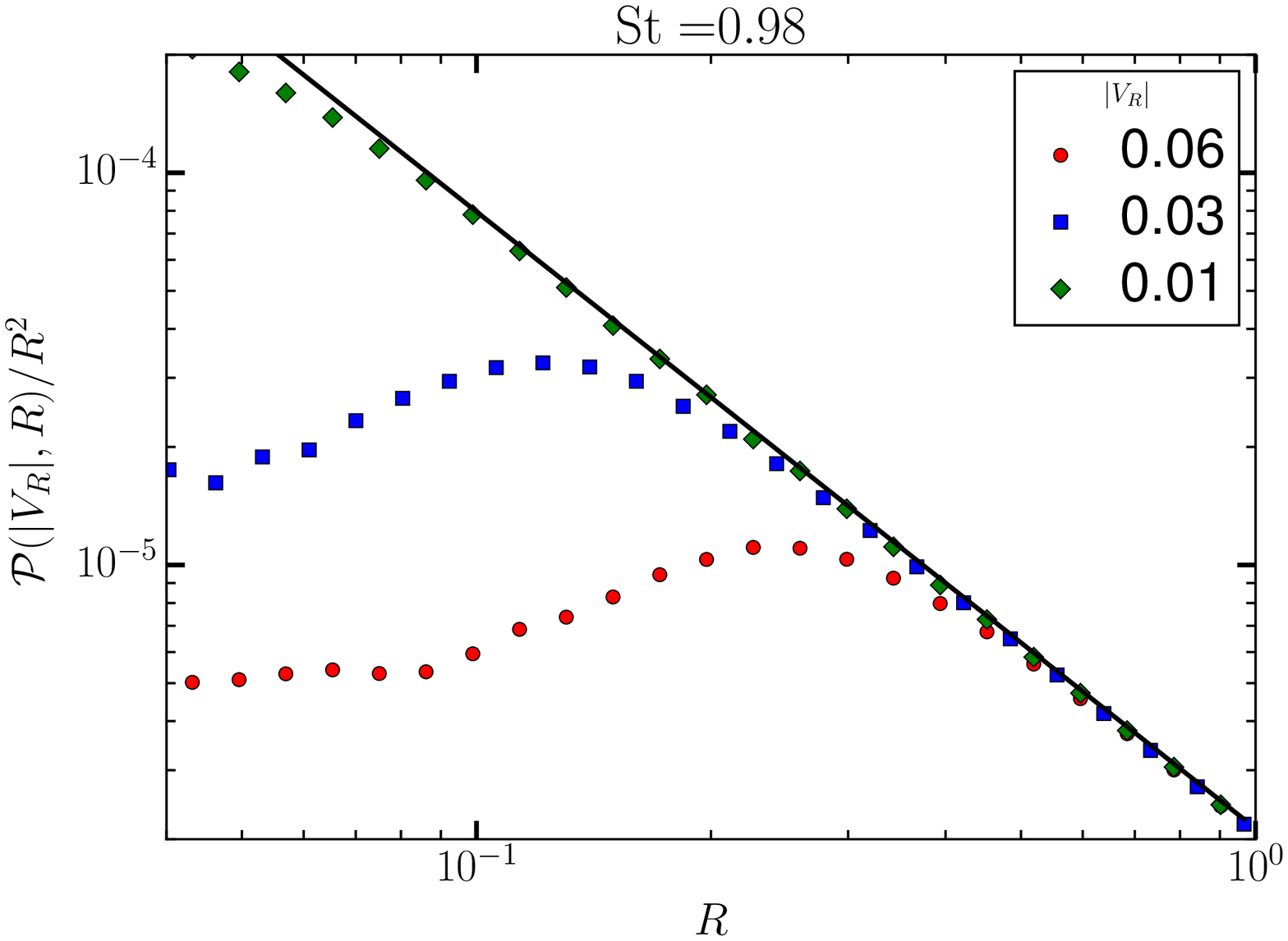} \\
\includegraphics[width=0.85\linewidth]{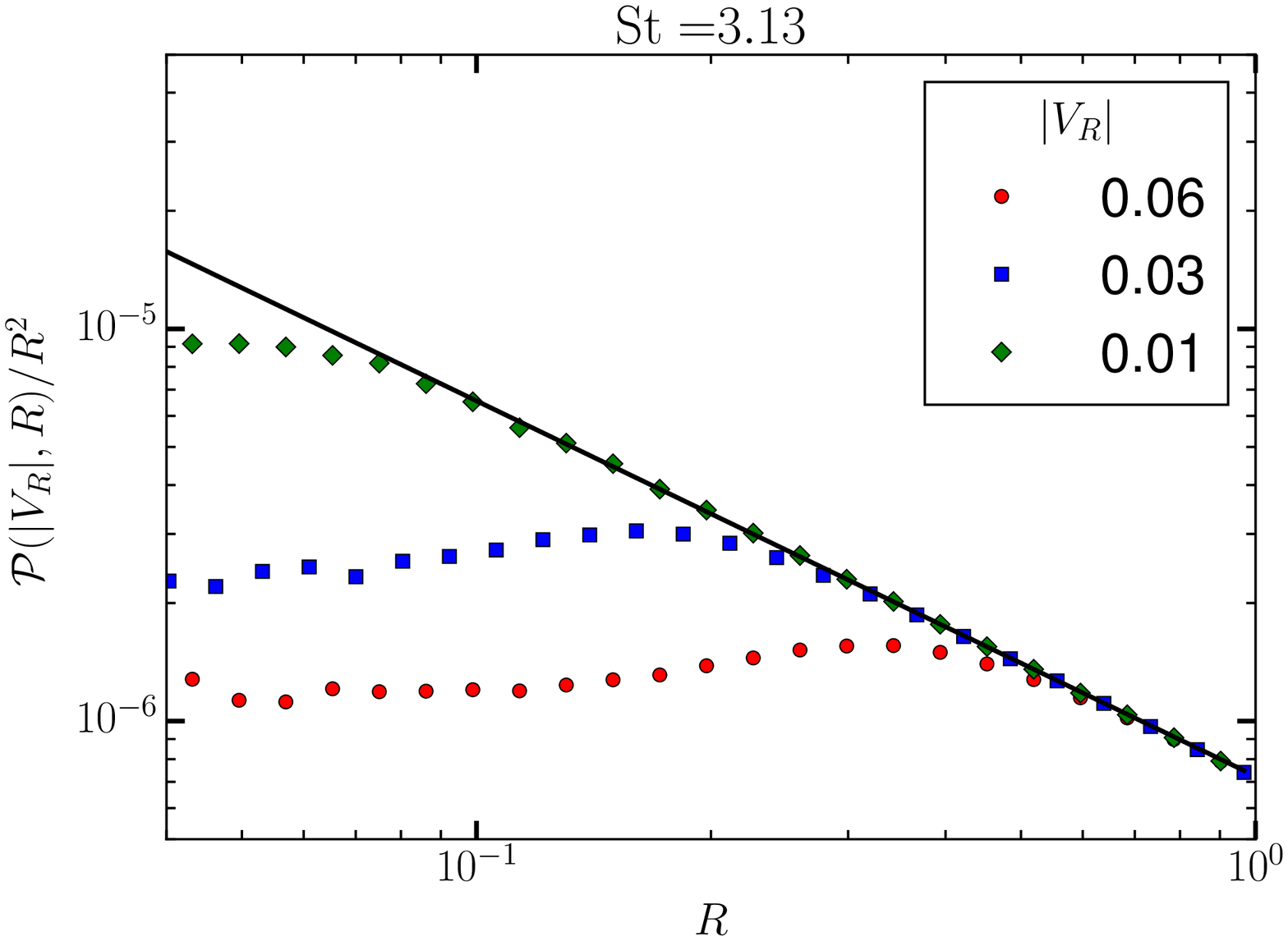}\\
\includegraphics[width=0.85\linewidth]{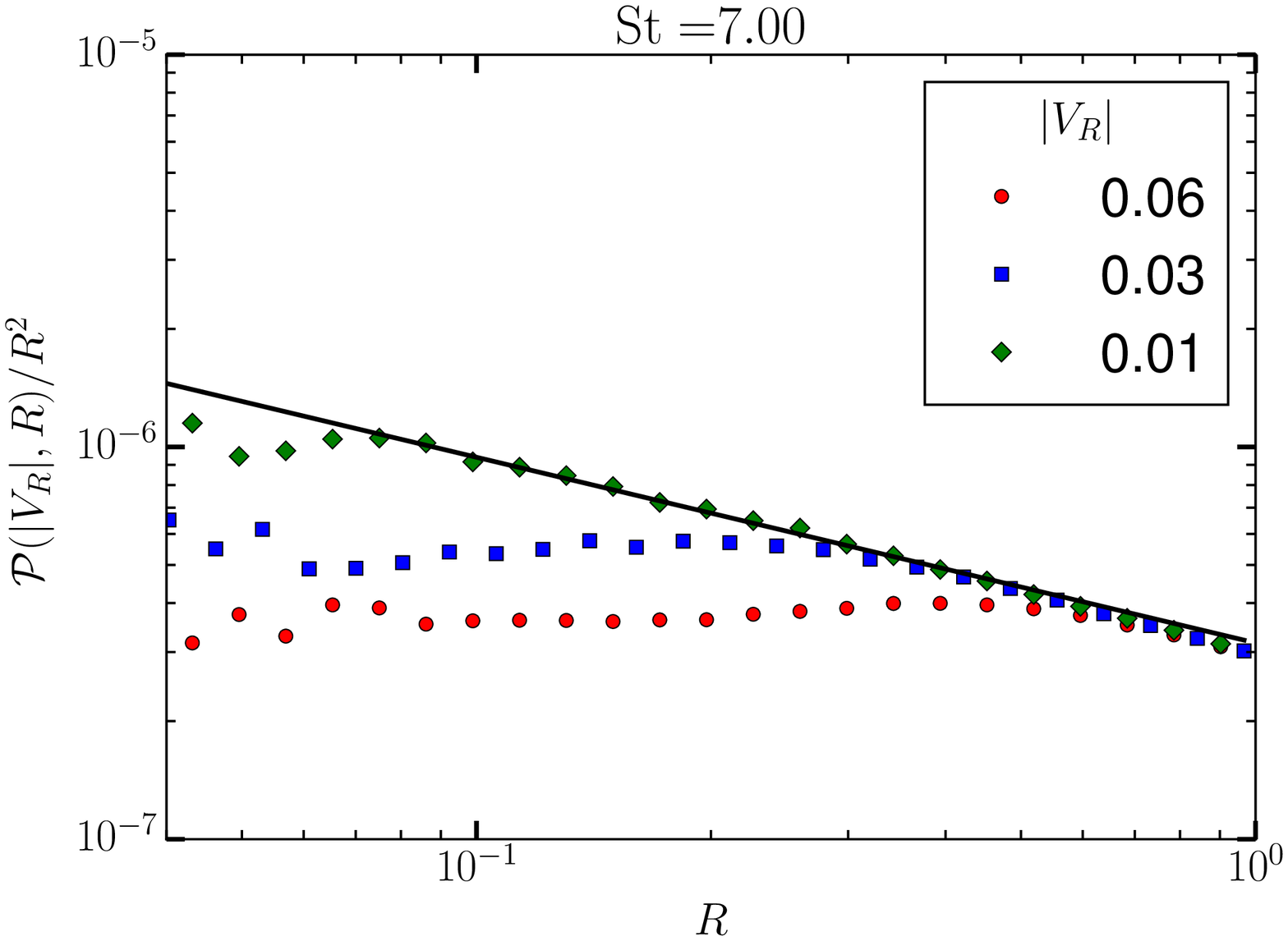}
\caption{(color online) log-log plots of $\mP(R,\VRa)/R^2$ versus $R$,
for four representative values of $\St$. For each value of
$\St$ we plot the curves for three fixed values of $\VRa$.
The solid black line in each plot
has slope $\Dtwo-4$, where $\Dtwo$ depends on $\St$
as shown in \Fig{fig:D2}}
\label{fig:pdf2}
\end{center}
\end{figure}
Continuing with our quantitative tests we plot in the four panels of
\Fig{fig:pdf2}, $\mP(R,\VRa)/R^2$ as a function of $R$ for four different
values of $\St$. On each panel we plot three different values of $\VRa$.
According to the theoretical prediction, \Eq{fscaling}, we expect to
find scaling with an exponent of $\Dtwo - 4$ which is indeed confirmed.

\begin{figure}
\begin{center}
\includegraphics[width=0.9\linewidth]{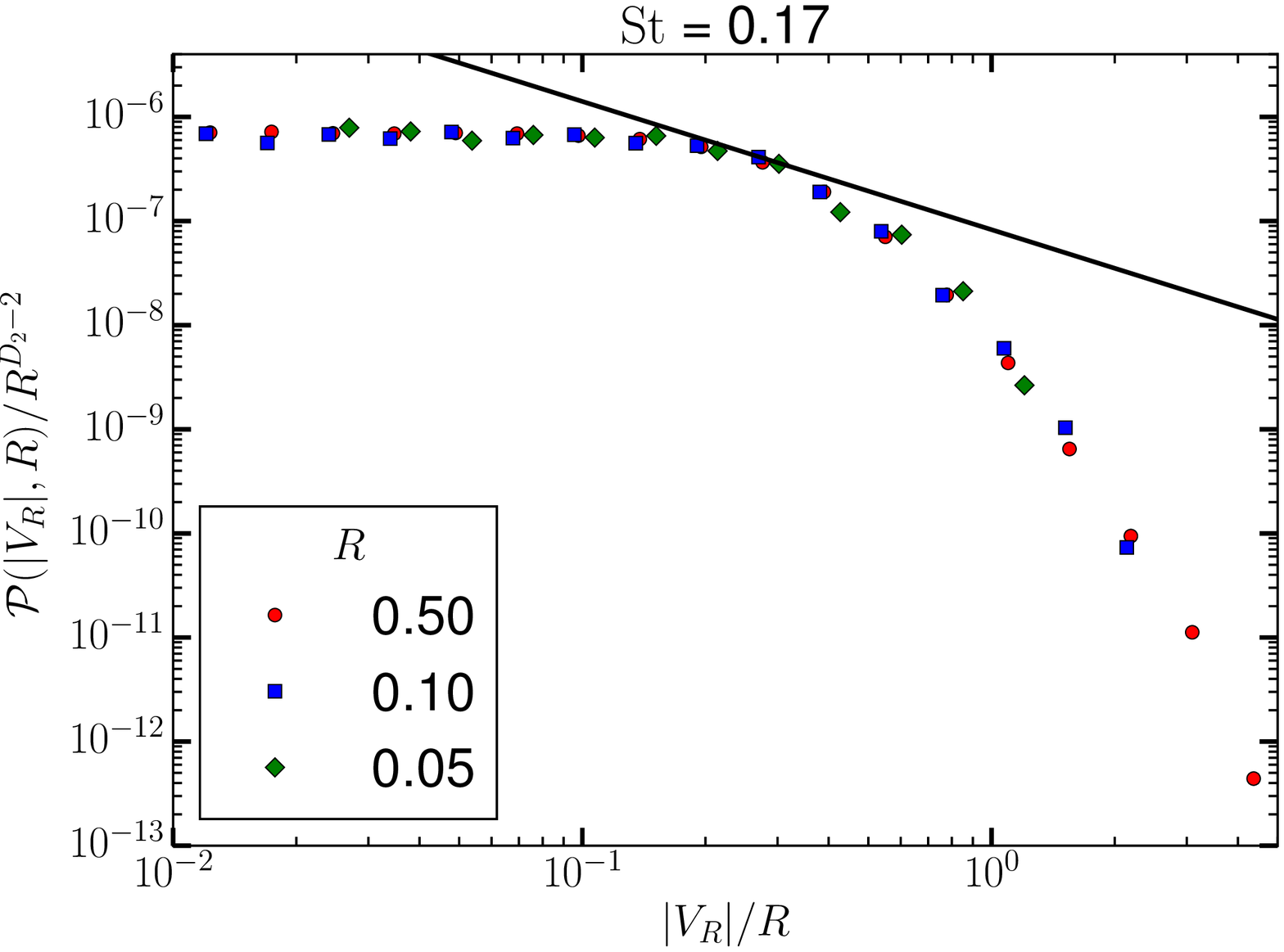}\\
\includegraphics[width=0.9\linewidth]{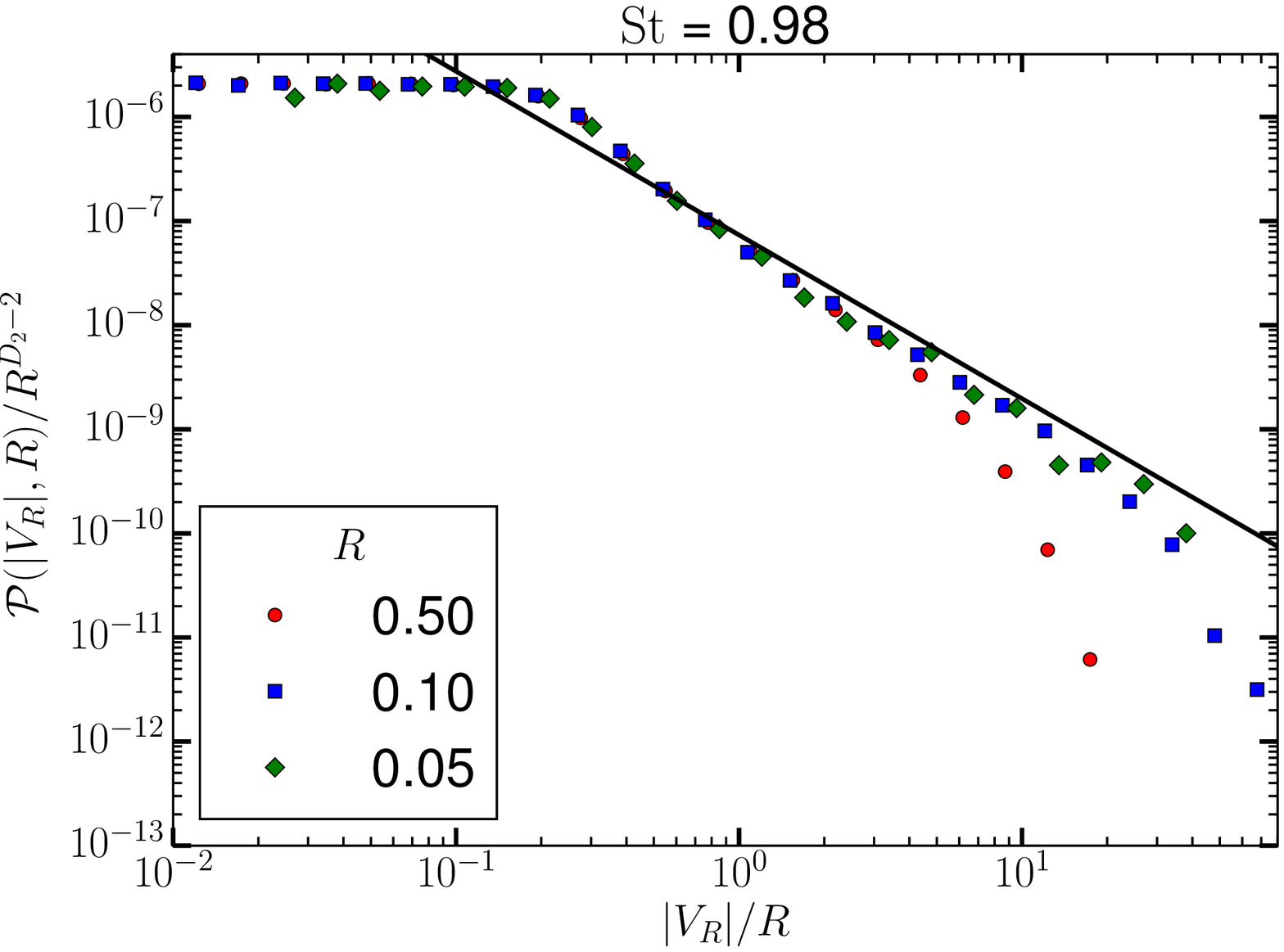} \\
\includegraphics[width=0.9\linewidth]{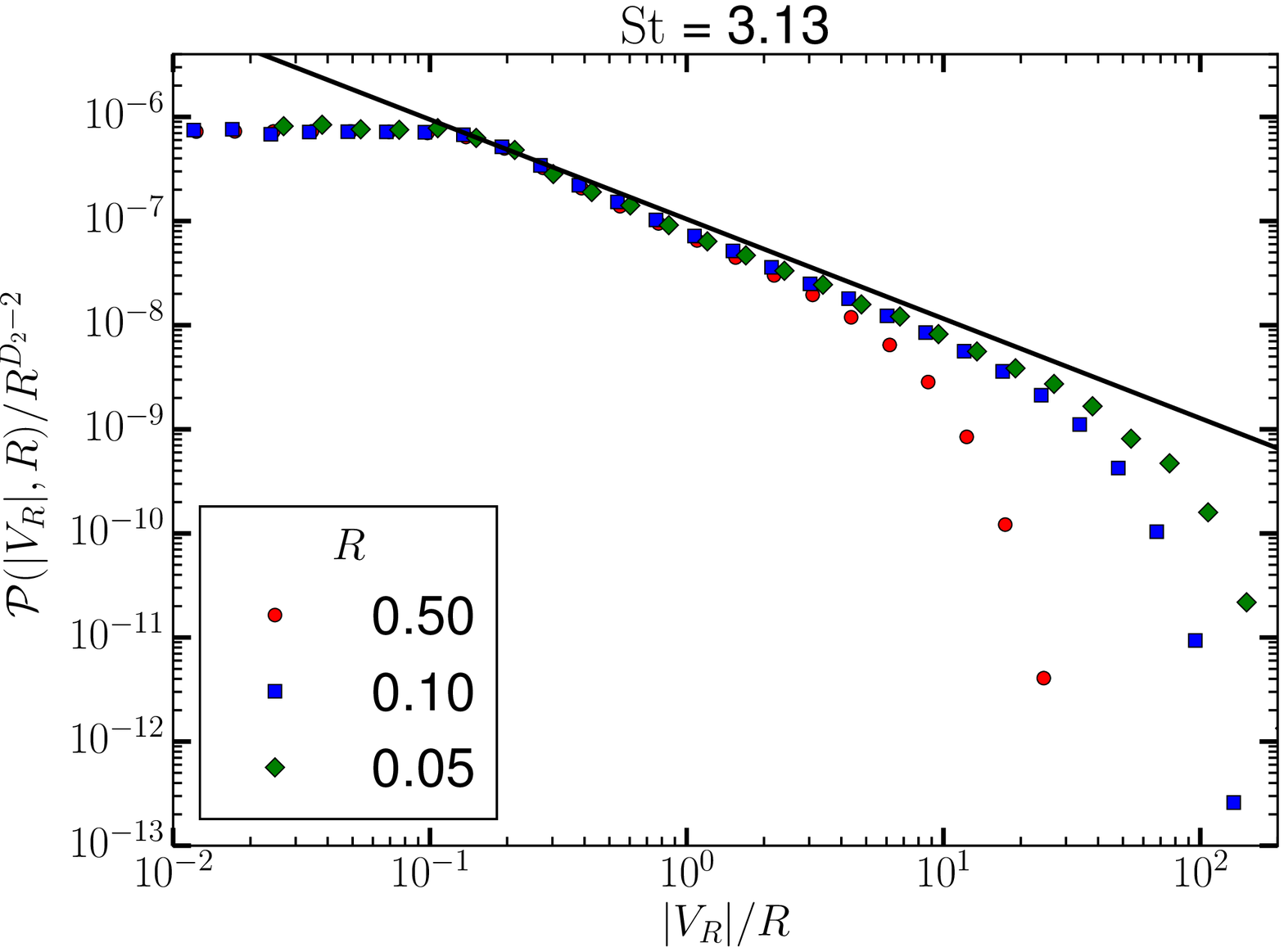}\\
\includegraphics[width=0.9\linewidth]{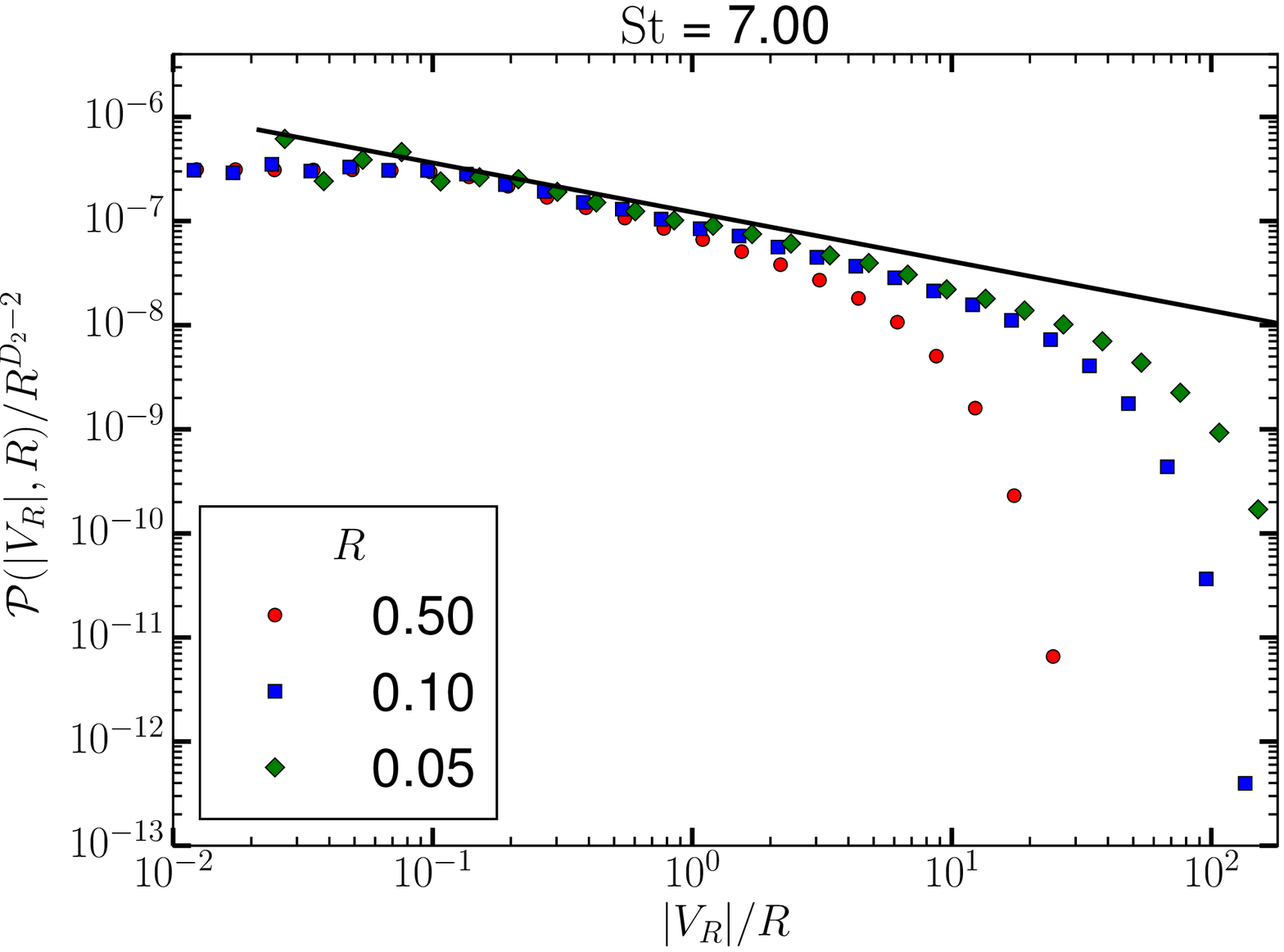}
\caption{(color online) log-log plots of $\rho(V_R,R)/R^{D_2-4}$ versus $V_R/R$,
for four representative values of $\St$. For each value of
$\St$ we plot the curves for three different values of $R$. The solid line in each plot
has slope $\Dtwo-4$, where $\Dtwo$ depends on $\St$.}
\label{fig:pdf1}
\end{center}
\end{figure}
Next let us consider the PDF of $\VRa$ for a
fixed $R$, $\PR(\VRa) = \mP(R,\VRa)$.
The scaling prediction for $\mP$ as given in \Eq{eq:fscaling} would imply
that $\mP(R,\VRa) \sim \VRa^{D_2-d-1}\sim \VRa^{D_2-4}$ for a fixed $R$.
Looking back at the contour plot of $\mP$ in \Fig{fig:jpdf} for, e.g.,
$\St=0.69$ we realize that $\PR$ can be obtained by taking a vertical
cut through this figure.
As the contour is almost horizontal until it reaches the matching line,
we expect $\PR$ to be independent of $\VRa$, for $\VRa < \zast R$ and
proportional to $R^{\Dtwo-2}$  until it reaches the upper cutoff
at $\VRa \approx\zast$.
Hence we expect that if we plot $\PR(\VRa)/R^{D_2-2}$ versus
$(\VRa/R)$ for different values of $R$ we would obtain a data
collapse to an universal function with a scaling exponent of $D_2-4$.
This we do for four different values of $\St$ in  four panels of
\Fig{fig:pdf1}. On each panel we plot the data for three different
values of $R<1$.
For the smallest $\St$, $\St = 0.17$, we do see a collapse of the data
but no scaling behavior.
This happens because the scaling appears due to the contribution
from the caustics and at small $\St$ we have not been able to probe
small enough distances $R$ to see scaling.
Indeed the scaling behavior appears for the next two values of $\St$,
$\St = 0.98$ and $3.13$ with the values of $D_2$ obtained from
\Fig{fig:D2}.
At large $\VRa$ we find departure from the data collapse.
This is what one would expect because the cutoff to the scaling
 behavior is set by $\zast$ which is independent of $R$.

Next we consider the scale $\zast$. The scaling theory has no way
of determining this, we calculate this as the best fit to our data and
plot it as a function of $\St$ in Appendix~\ref{app:zstar}, \Fig{fig:zstar}.
We find that $\zast$ varies very little -- within a factor of two --
as $\St$ changes by two orders of magnitudes.
In principle $\zast$ can also depend on $\Rey$, within the range
of $\Rey$ studied in our simulations, it does not have any
significant dependence on $\Rey$.

\begin{figure}
\begin{center}
\includegraphics[width=0.9\linewidth]{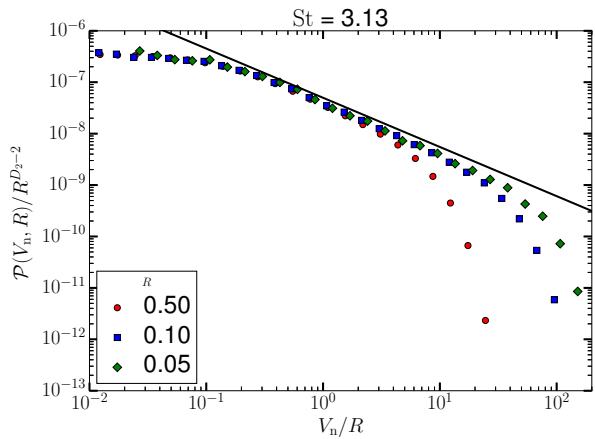}
\caption{(color online) log-log plot of  $\mP(R,\VRn)/R^{D_2-2}$ versus
  $\VRn/R$, for $\St=3.13$ and three values of $R$. Here $\VRn$ is the
  negative relative longitudinal velocity between two particles
  separated by a distance $R$. The solid black line has slope $D_2 - 4$. 
  Comparing this figure with \Fig{fig:pdf1} we find that the
distributions of $\VRn$ and $\VRa$ has the same scaling behavior.}
\label{fig:vrn}
\end{center}
\end{figure}
Note finally that so far we have presented all our numerical results
for the absolute value of $\VR$ ignoring its sign.
But the sign of $\VR$ is crucial as it sets the collision kernel.
We have checked that the statistics of negative side of $\VR$, defined by
$\VRn$, follows exactly the same scaling behavior as that of $\VRa$.
To demonstrate this we plot in \Fig{fig:vrn} the scaling collapse of
the joint PDF of $R$ and $\VRn$ for $\St = 3.13$.
Comparing this figure with \Fig{fig:pdf1} we find that the
distributions of $\VRn$ and $\VRa$ has the same scaling behavior.
\section{Conclusions}
\label{conc}
To summarize, the joint PDF, $\mP$, gives a complete description of the statistics of relative
velocities and distances of a pair of heavy inertial particles in homogeneous and
turbulent flows.
Quantities of more practical interest, for example, the spatial
clustering and the collision kernel follows from this joint PDF.
Our simulations confirms the theoretically predicted asymptotic
mirror symmetry of  $\mP$ about a matching line $\Upsilon(R)=\zast R$.
Furthermore, our DNS confirms the scaling behavior predicted in
Ref.~\cite{gustavsson2011distribution}.
The scale $\zast$ sets the cutoff to the scaling behavior of
PDF of $\VRa$, hence it sets the maximum possible relative velocities; as we
have non-dimensionalized all velocities using $\ueta$. This implies
that the maximum collision speed of heavy inertial particles in
turbulent flows is of the order of $\zast\ueta$.
In our DNS we find that $\zast$ depends very weakly on $\St$ in
contradiction to  the results of white-noise model in
Ref~\cite{gustavsson2011distribution}, who found $\zast \sim \St$.

To the best of our knowledge, the joint PDF and its scaling properties have
never been calculated from DNS of homogeneous and isotropic turbulent
flows, although it has been calculated numerically for
two-dimensional random smooth flows and analytically for a
one-dimensional white-noise model~\cite{gustavsson2011distribution}.
Other than Ref.~\cite{gustavsson2011distribution},
Ref.~\cite{pan2013turbulence} has presented theoretical arguments and
DNS results on PDF of $\VRa$ for a given $R$.
However Ref.~\cite{pan2013turbulence} did not elucidate the scaling
nature of the PDF instead concentrated on the large-$\VRa$ cutoff of
the PDF.
Ref.~\cite{per+jon15} have obtained PDF of $\VR$ for several values of
$R$ from DNS and had confirmed the scaling behavior for $\St \approx
1$, but they did not study the asymptotic mirror symmetry.
There are also limited experimental
data~\cite{de2010measurement,saw+bew+bod+ray+bec14} available for $\VR$
but most
of it is limited to $R \simeq \eta$ hence is not sufficient to study
the scaling behavior as we do.

\section{Acknowledgment}
This work is supported by the grant Bottlenecks
for particle growth in turbulent aerosols from
the Knut and Alice Wallenberg Foundation (Dnr. KAW
2014.0048).  The computations
were performed on resources provided by the
Swedish National Infrastructure for Computing (SNIC)
at PDC.

\bibliographystyle{prsty}
\bibliography{ref,turb_ref}
\appendix
\section{$\dive \uu$ along the particle tracks}
\label{app:divu}

\begin{figure}
\begin{center}
\includegraphics[width=0.9\linewidth]{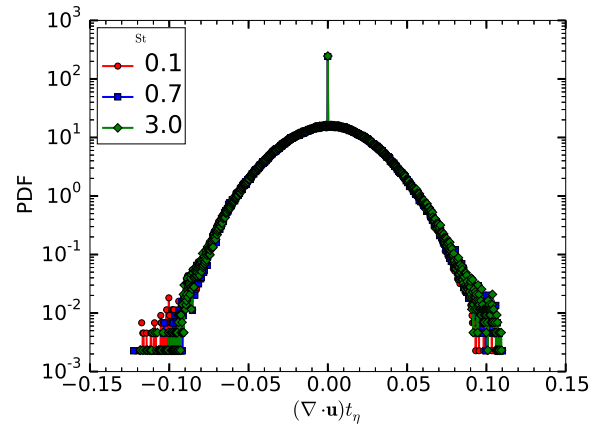}
\caption{(color online) PDFs of $\nabla \cdot \uu$ calculated along the trajectories of
particles for three values of $\St$, from the run {\tt R2}}
\label{fig:divu}
\end{center}
\end{figure}

In our simulations the flow velocity field $\uu$ is weakly compressible. To test the effect of
this weak compressibility on the dynamics of heavy inertial particles, we calculate $\dive
\uu$ at the position of particles. This is done by first calculating $\dive \uu$ at
the grid points and then interpolating it to the positions of particles. We plot the
PDFs of $\dive \uu$ along the trajectories of particles having three different values of
$\St$ in \Fig{fig:divu}. We non-dimensionalize $\dive \uu$ by the inverse of $\teta$, which
gives an idea of shear at the Kolmogorov scale $\eta$. In \Fig{fig:divu}, we see that the
PDFs are Gaussian with a high peak at zero and have a small width compared to $1/\teta$.
We also observe that these distributions does not depend on the value of $\St$. These
observations suggest that $\dive \uu$ does not attain very high values along the
trajectories of particles and effects of compressibility are weak.

\section{Matching scale $\zstar$}
\label{app:zstar}

\begin{figure}
\begin{center}
\includegraphics[width=0.9\linewidth]{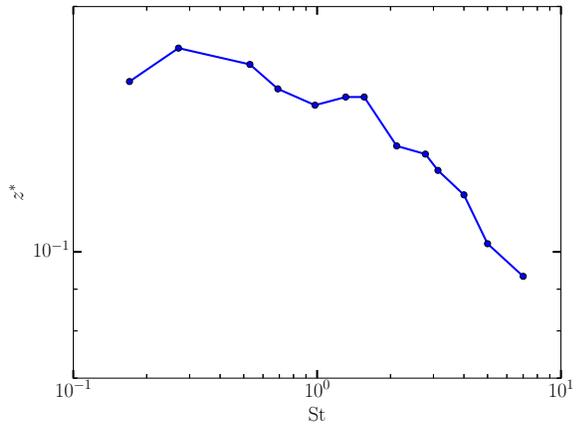}
\caption{(color online) Matching scale $\zast$ as a function of $\St$
  from the run {\tt R2}}
\label{fig:zstar}
\end{center}
\end{figure}

We show in \Fig{fig:jpdf} that the joint PDFs in regions \I ~ and \II ~ can be matched along a
straight line $\VRa = \zstar R$, for $R < 1$. We find the matching scale $\zstar$ by fitting a
straight line through the points where contour lines in \Fig{fig:jpdf} turns from vertical
to horizontal. We show the plot of $\zstar$ as a function of $\St$ if \Fig{fig:zstar}. We
find that $\zstar$ weakly depends on $\St$. Its value remains close to $10^{-1}$. Values
of $\zstar$ for small values of $\St$ are not reliable because for small $\St$ [c.f. \Fig{fig:jpdf} (a)] we do not see the asymptotic regime in region \II .

\end{document}